\documentclass[reprint,amsmath,amssymb,aps,footinbib]{revtex4-1}  % 2-column
\usepackage{dcolumn}
\usepackage{color}
\usepackage{bm}
\usepackage{xspace}
\usepackage{graphicx} % arXiv

\newcommand{\be}{\begin{equation}}
\newcommand{\ee}{\end{equation}}
\newcommand{\bea}{\begin{eqnarray}}
\newcommand{\eea}{\end{eqnarray}}
\newcommand{\beaa}{\begin{eqnarray*}}
\newcommand{\eeaa}{\end{eqnarray*}}
\newcommand{\alphaI}{{$\alpha$-(BEDT-TTF)$_2$I$_3$}\xspace}

\begin{document}

\title{
  Evidence for three-dimensional Dirac semimetal state
  in strongly correlated organic quasi-two-dimensional material
}

\author{Naoya Tajima}
\email{naoya.tajima@sci.toho-u.ac.jp}
\affiliation{Department of Physics, Toho University, Funabashi, Chiba 274-8510, Japan}
\author{Takao Morinari}
\email{morinari.takao.5s@kyoto-u.ac.jp}
\affiliation{Course of Studies on Materials Science,
  Graduate School of Human and Environmental Studies,
  Kyoto University, Kyoto 606-8501, Japan
}
\author{Yoshitaka Kawasugi}
\affiliation{Department of Physics, Toho University, Funabashi, Chiba 274-8510, Japan}
\author{Ryuhei Oka}
\affiliation{Department of Chemistry, Ehime University, Matsuyama 790-8577, Japan}
\author{Toshio Naito}
\affiliation{Department of Chemistry, Ehime University, Matsuyama 790-8577, Japan}
\author{Reizo Kato}
\affiliation{RIKEN, Hirosawa 2-1, Wako-shi, Saitama 351-0198, Japan}

\date{\today}

\begin{abstract}
  The three-dimensional Dirac semimetal is distinct
  from its two-dimensional counterpart due to its dimensionality and symmetry.
  Here, we observe that molecule-based
  quasi-two-dimensional Dirac fermion system, \alphaI,
  exhibits chiral anomaly-induced negative magnetoresistance
  and planar Hall effect
  upon entering the coherent inter-layer tunneling regime under high pressure.
  Time-reversal symmetry is broken due to the strong electronic correlation effect,
  while the spin-orbit coupling	effect is negligible.
  The system provides an ideal platform for investigating
  the chiral anomaly physics by controlling dimensionality
  and strong electronic correlation.
\end{abstract}

\maketitle

%\section{Introduction}
Spatial dimensionality plays a crucial role in the research of Dirac semimetal physics. 
In a two-dimensional (2D) system, the system exhibits
an unconventional half-integer quantum Hall effect\cite{Novoselov2005,Zhang2005}
when the Dirac fermions are massless.
An instability of creating a mass gap leads to
a remarkable state of matter,
2D topological insulator\cite{KaneMele2005a,KaneMele2005b,Hasan2010,Qi2011}.
While in a three-dimensional (3D) system,
the system exhibits intriguing transport properties
associated with the chiral anomaly\cite{Nielsen1983,Son2013,Burkov2015,Nandy2017,Burkov2017,Xiong2015,Huang2015a,Hirschberger2016,Zhang2016,Li2016}.

To realize a Weyl semimetal, it is necessary to break 
either time-reversal symmetry or inversion symmetry, or both. 
On the other hand, a Dirac semimetal can be realized 
even when both time-reversal and inversion symmetries are preserved.

%----------------------------------------------------------------------------------
From the quantum field theory, it is well known	that
the chiral anomaly exists in 3D but not in 2D\cite{Peskin1995}.
Usually, Dirac semimetals are divided into 2D or 3D Dirac semimetals,
and no candidate material has connected them so far.
In this Letter, we report that \alphaI is such a system:
Under high pressure it is a 2D Dirac semimetal
at high temperatures and a 3D Dirac semimetal at low temperatures.

\alphaI has been studied extensively\cite{Kajita2014}
as a quasi-2D Dirac semimetal\cite{Katayama2006,Tajima2000,Tajima2006}.
(Here, BEDT-TTF is bis(ethylenedithio)tetrathiafulvalene.)
A single crystal of \alphaI with the space group of $P\overline{1}$
has a quasi-two-dimensional structure
consisting of conductive layers
of BEDT-TTF molecules and insulating layers of I$_3^-$ anions\cite{Bender1984}.
The unit cell consists of four BEDT-TTF molecules,
A, A$^{\prime}$, B, and C in the conduction layer\cite{Mori1984}.
% charge order
Due to the strong electronic correlation,
\alphaI is a Mott insulator with charge order under ambient pressure.
The system undergoes a metal-insulator
transition\cite{Kartsovnik85,Schwenk1985,Kajita1992}
at 135~K with forming a stripe pattern of charge order
as confirmed by $^{13}$C-NMR measurement\cite{Takano2001,Takahashi2003}.
A and A$^{\prime}$ molecules and B and C molecules stack along $a$ axis,
and alternating pattern of charge $e$ and 0 is formed,
and the system exhibits a charge stripe pattern along $b$ axis.
Theoretical analysis\cite{Seo2000,Kino1996} revealed that
the short-range inter-site Coulomb repulsion plays an important role,
and the result is consistent with the experiment\cite{Takahashi2003}.
Above 1.5 GPa the charge ordered insulating state
is unstable\cite{Tajima2000,Tajima2006},
and the system enters
the quasi-2D massless Dirac fermion phase\cite{Kajita2014}.
According to the band calculation of this material under high pressure\cite{Katayama2006},
the electronic structure is described by massless Dirac fermions,
and this result is confirmed
by the first principles calculations\cite{Ishibashi2006,Kino2006}.
The first band and the second band contacts each other
at two points in the first Brillouin zone,
and the system exhibits the linear energy dispersion.

Since the system has only inversion and time-reversal symmetries,
the Dirac nodes are not at high-symmetric points in the first Brillouin zone.
A salient feature is that the Dirac point is precisely at the Fermi energy.
This has been confirmed experimentally
by the observation of the negative inter-layer magnetoresistance
where the zero-energy Landau level at the Fermi level
plays an essential role\cite{Osada2008,Tajima2009,Morinari2009,Goerbig2008}.
The phase of the Dirac fermions is confirmed from 
the Shubnikov-de Haas oscillation of the hole doped sample,
where the sample is placed on polyethylene naphthalate substrate\cite{Tajima2013}.
A breakdown of the Korringa law suggests that the system
is in a strong coupling regime\cite{Hirata2017}.
When the system approaches the quantum critical point (QCP)
between the massless Dirac fermion phase and charge ordering phase,
the Fermi velocity decreases without creating the mass gap\cite{Unozawa2020}.
The decrease of the Fermi velocity is associated with
the strong on-site Coulomb repulsion\cite{Tang2018}.

One of the present authors proposed\cite{Morinari2020} that,
upon entering a coherent interlayer tunneling regime,
the system
becomes a 3D Dirac semimetal phase\cite{Young2012a,Wang2012,Yang2014}.
The inter-site short-range Coulomb repulsion introduces phase modifications
in the hopping parameters that break time-reversal and inversion symmetries.
In addition, the interlayer tunneling between different molecules makes
the system into the 3D type-II Dirac semimetal\cite{Soluyanov2015}. 
%
%----------------------------------------------------------------------------------

  Time-reversal symmetry is broken at high temperatures\cite{Morinari2023flux} because its energy scale is associated with the short-range Coulomb repulsion, which is on the order of 0.1~eV.
  Due to the broken inversion symmetry, the positions of the Dirac points are not related by this symmetry. For example, from the mean field calculation at 0.8 GPa\cite{Morinari2020}, we find two Dirac points at $(0.939,-0.806, \pm 1.571)$ and another two at
  $(-0.908,0.760, \pm 1.571)$.

%----------------------------------------------------------------------------------
Recently we observed a peak structure at low temperatures
in the interlayer magnetoresistance when a magnetic field is
parallel to the plane\cite{Tajima2023}
that suggests coherence in the interlayer tunneling\cite{Hanasaki1998}.
We also examined the effect of spin-orbit coupling and found
that it does not affect the electronic structure\cite{Morinari2023LT29}.

\begin{figure}[htbp]
  \includegraphics[width=0.8 \linewidth]{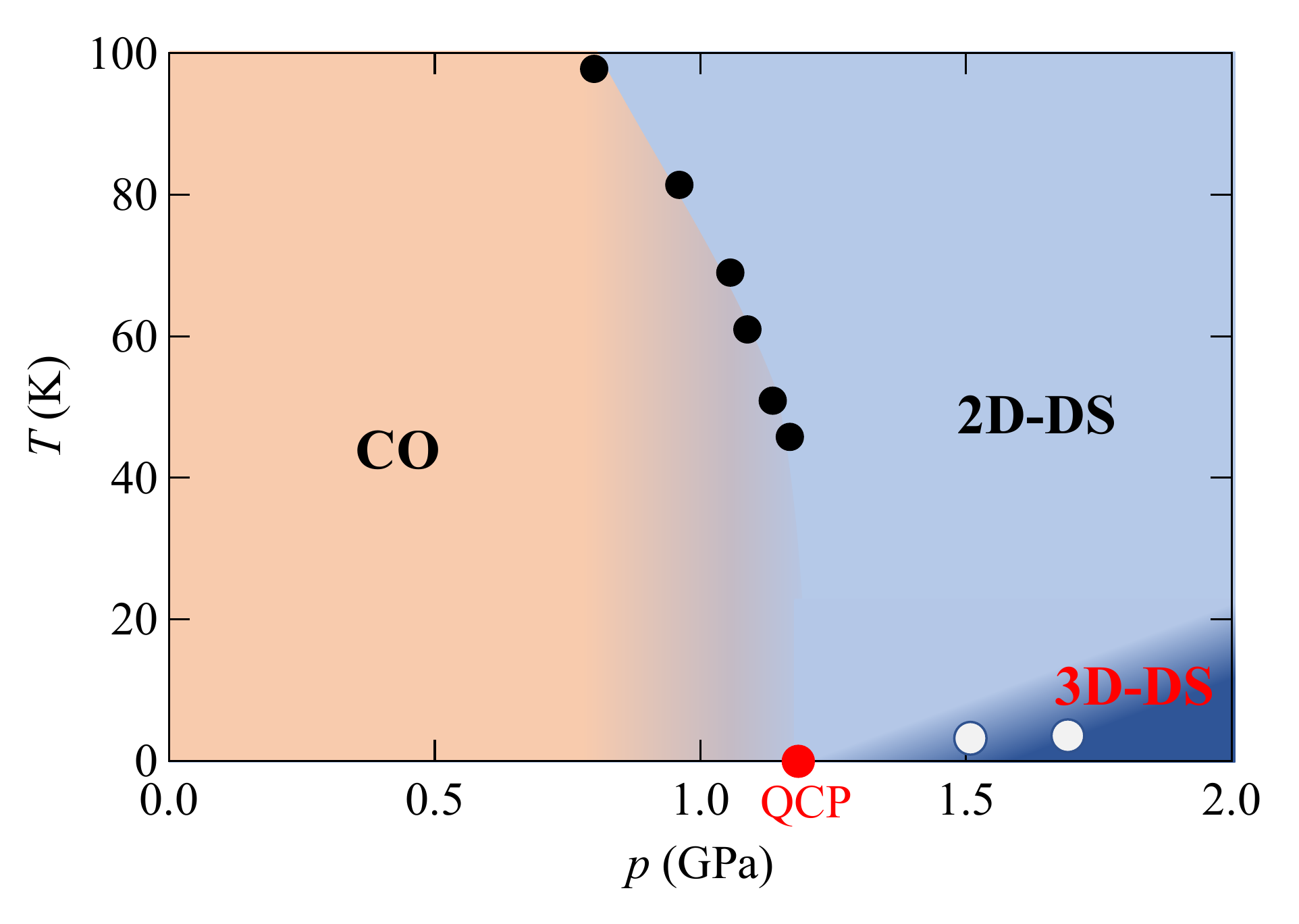}
  \caption{
    \label{fig:phase_diagram}
    Pressure-temperature phase diagram of \alphaI. 
    The 3D Dirac semimetal (3D-DS) was identified based on the transport measurements
    at the two points indicated by the white circles.
    Note that the boundary between 2D Dirac semimetal (2D-DS) and 3D-DS is unclear.
    The charge ordered state is denoted by CO.
  }
\end{figure}

In this Letter, we report on the evidence
that the 3D Dirac semimetal is realized in \alphaI.
We clearly observe the negative magnetoresistance and the planar Hall effect
that are associated with the chiral anomaly.
To our knowledge, this is the first observation of the chiral anomaly effect
in organic conductors.
Combined with the previous results\cite{Unozawa2020},
our phase diagram is shown in Fig.~\ref{fig:phase_diagram}.

%-------------------------------------------------------------------
% Figure: phase diagram
%-------------------------------------------------------------------
%-------------------------------------------------------------------

%{\it Expeiments.-}\\
A sample on which six electrical leads are 
attached is put in a Teflon capsule 
filled with the pressure medium (Idemitsu DN-oil 7373) 
and then the capsule is set in a clamp type pressure cell 
made of hard alloy MP35N.
We measured resistance
by a conventional DC method with an electrical current $I$ of 10 $\mu$A 
which was applied along the $a$-axis for four samples.
Under the pressure of 1.7 GPa,
we measured the magnetic field dependence of the magnetoresistance 
$\Delta \rho / \rho_0 = \{ \rho_{xx}(B)-\rho_{xx}(0)\} / \rho_{xx}(0)$ 
for three samples. 
On the other hand,  
the magnetoresistance and the planar Hall resistivity 
$\rho_{xy}^{ \rm PHE}$ were investigated 
as functions of the magnetic field 
and the azimuthal angle $\phi$ to the $a$-axis in the $ab$ plane 
for another sample at 1.5 GPa.

\begin{figure}[htbp]
  \includegraphics[width=0.9 \linewidth]{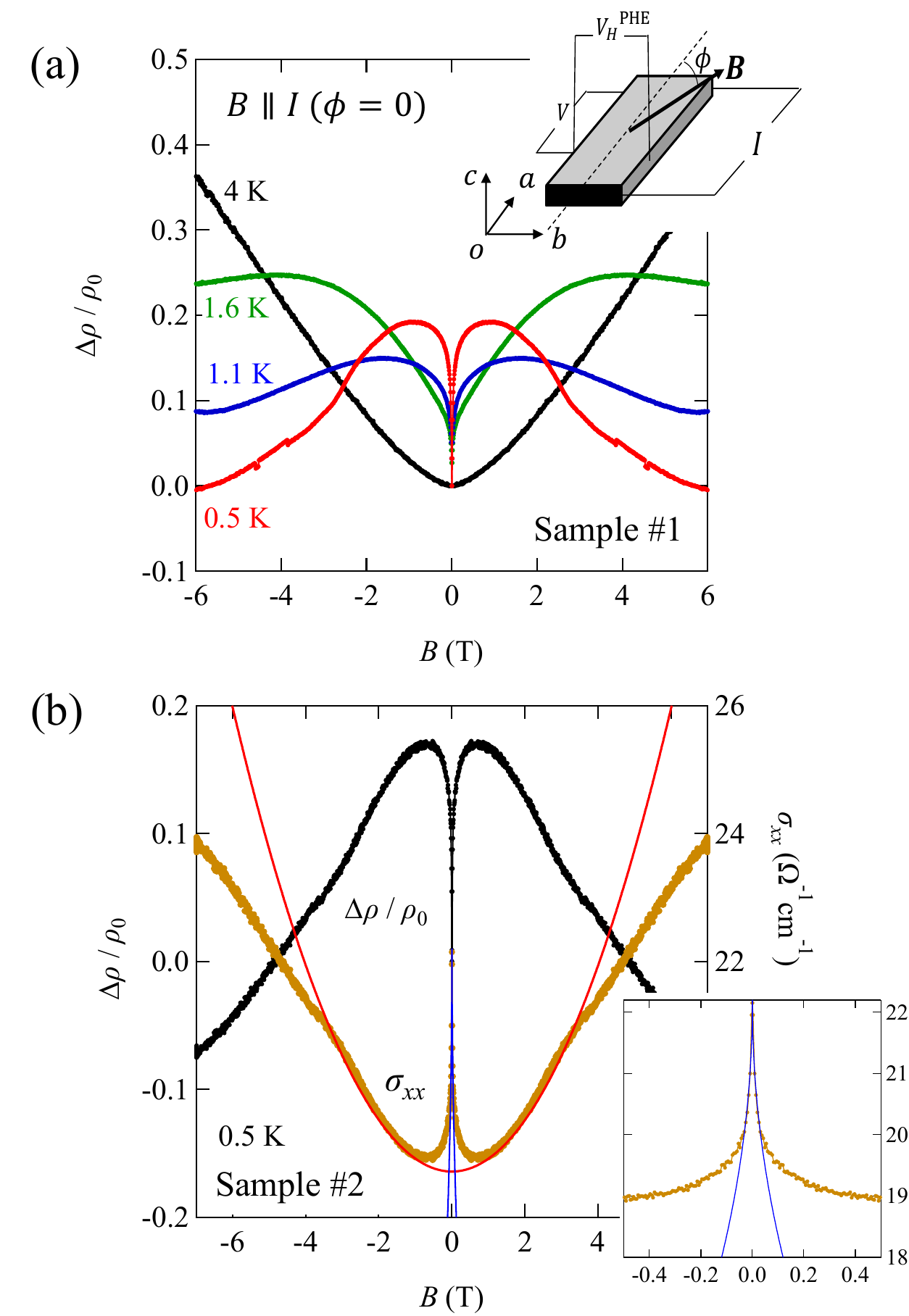}
  \caption{
    \label{fig:nmr_exp}
    (a) Magnetic field dependence of the longitudinal magnetoresistance
    at temperature below 4 K for Sample \#1.
    The inset shows the measurement configurations
    for the magnetoresistance and the planar Hall effect.
    (b) Magnetic field dependence of the longitudinal magnetoresistance and the conductivity at 0.5 K for Sample \#2. The red curve is the equation $\sigma_{xx}=\sigma_{0}+C_{ca} B^2$
with $\sigma_{0}~22.1$ $\Omega^{-1} {\rm cm}^{-1}$ and 
$C_{ca}\sim 0.2$ $\Omega^{-1} {\rm cm}^{-1} T^{-2}$.
The inset shows the conductivity near zero field
which is proportional to $\sqrt{|B|}$, indicated by the blue line. 
}
\end{figure}

Figure~\ref{fig:nmr_exp}(a) shows magnetic field dependence of 
the longitudinal magnetoresistance for Sample \#1 at temperature below 4 K. 
The magnetoresistance behaves as 
$\Delta \rho / \rho_0 \propto |B|^{\alpha}$ with $\alpha \sim 1.8$ at 4 K.
This 2D Dirac semimetal behavior changes dramatically at low temperatures.
At temperature below 1.6 K, 
negative magnetoresistance was observed.
The result demonstrates the presence of chiral anomaly.
The lower the temperature, the clearer the negative magnetoresistance.
Another sample (Sample \#2) also showed a negative magnetoresistance 
in a magnetic field parallel to the electric field at 0.5 K 
as shown in Fig.~\ref{fig:nmr_exp}(b).
Here, the conductivity $\sigma_{xx}=1/\rho_{xx}$ is expressed as
$\sigma_{xx}=\sigma_{0}+C_{ca} B^2$.
From the fit to the data in $-3$ T $< B <$ $3$ T
we find $\sigma_{0} \simeq 22.1$ $\Omega^{-1} {\rm cm}^{-1}$
and $C_{ca} \sim 0.2$ $\Omega^{-1} {\rm cm}^{-1} {\rm T}^{-2}$.
At low fields the conductivity scales as $\sqrt{|B|}$
that is associated with weak anti-localization\cite{Lu2015}
as shown in the inset of Fig.~\ref{fig:nmr_exp}(b).
Meanwhile, $\sigma_{xx}$ deviates
from the $\sigma_{xx} \propto B^2$ law for $|B|>3$ T.
Similar behaviors were also observed
in other Dirac/Weyl semimetals\cite{Huang2015a,Xiong2015,Zhang2016}.
%-------------------------------------------------------------------
% Figure:
%-------------------------------------------------------------------
%-------------------------------------------------------------------

\begin{figure}[htbp]
  \includegraphics[width=0.85 \linewidth]{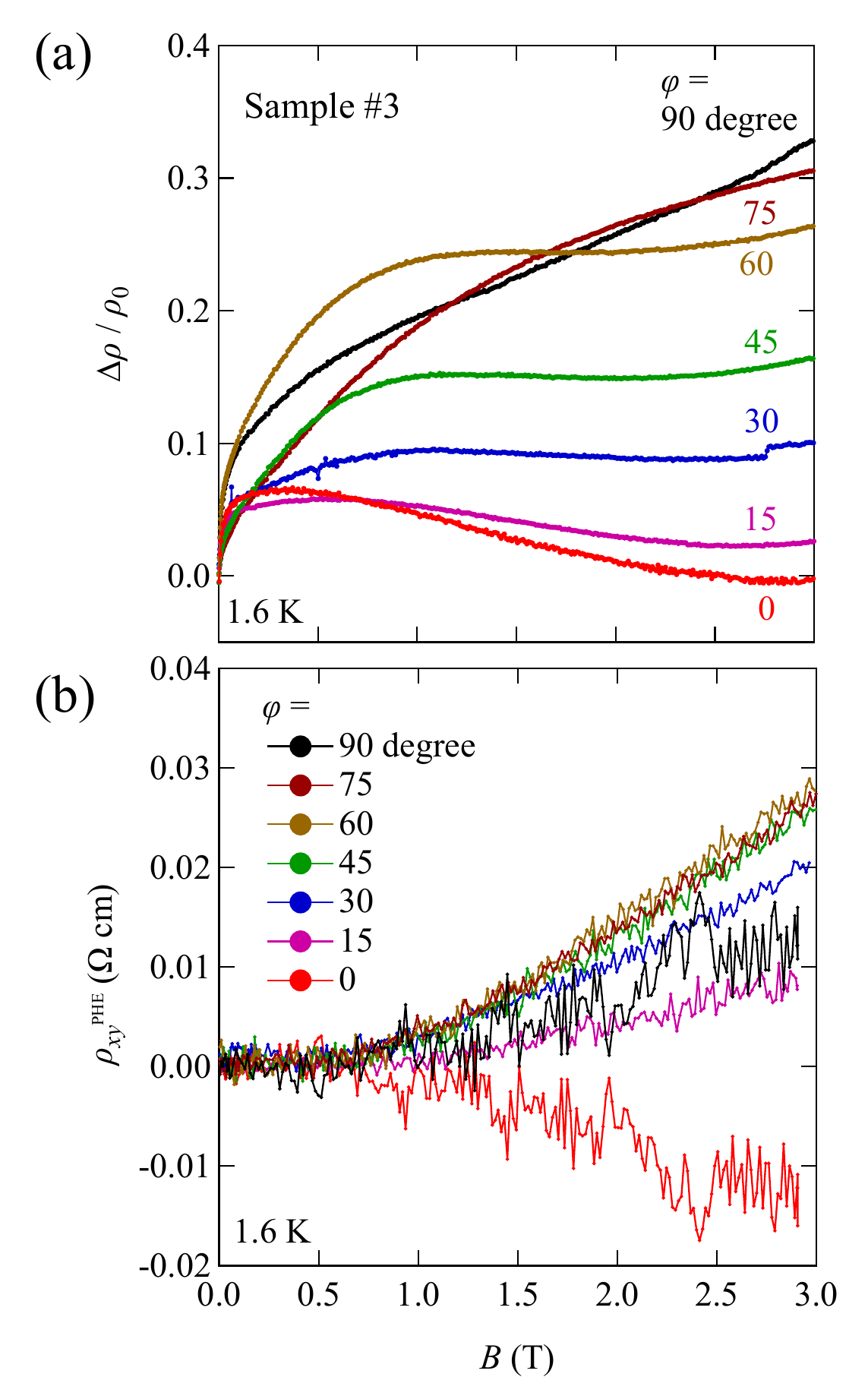}
  \caption{
    \label{fig:phe_exp}
    Magnetic field dependence of the magnetoresistance (a)
    and the planar Hall resistance (b) at 1.6 K for Sample \#3.
      We measured them by changing
      the azimuthal angle $\phi$ by 15 degrees.
   }
\end{figure}

% planar Hall effect
Figure~\ref{fig:phe_exp} shows
the planar Hall effect and the azimuthal angle dependence of
the negative magnetoresistance.
We measured the magnetoresistance and
the planar Hall resistivity $\rho_{xy}^{\rm PHE}$ of Sample \#3
at 1.6 K as shown in Fig.~\ref{fig:phe_exp}(a).
The negative magnetoresistance was observed for $\phi < 15^{\circ}$
while it became positive for $\phi > 15^{\circ}$.
Each planar Hall resistivity, on the other hand, is not linear but depends on $B^2$.
The non-zero planar Hall resistivity at $\phi=0$ could originate from
slight differences in the electric current and the magnetic field directions.
%-------------------------------------------------------------------
% Figure:
%-------------------------------------------------------------------
%-------------------------------------------------------------------

\begin{figure}[htbp]
  \includegraphics[width=0.85 \linewidth]{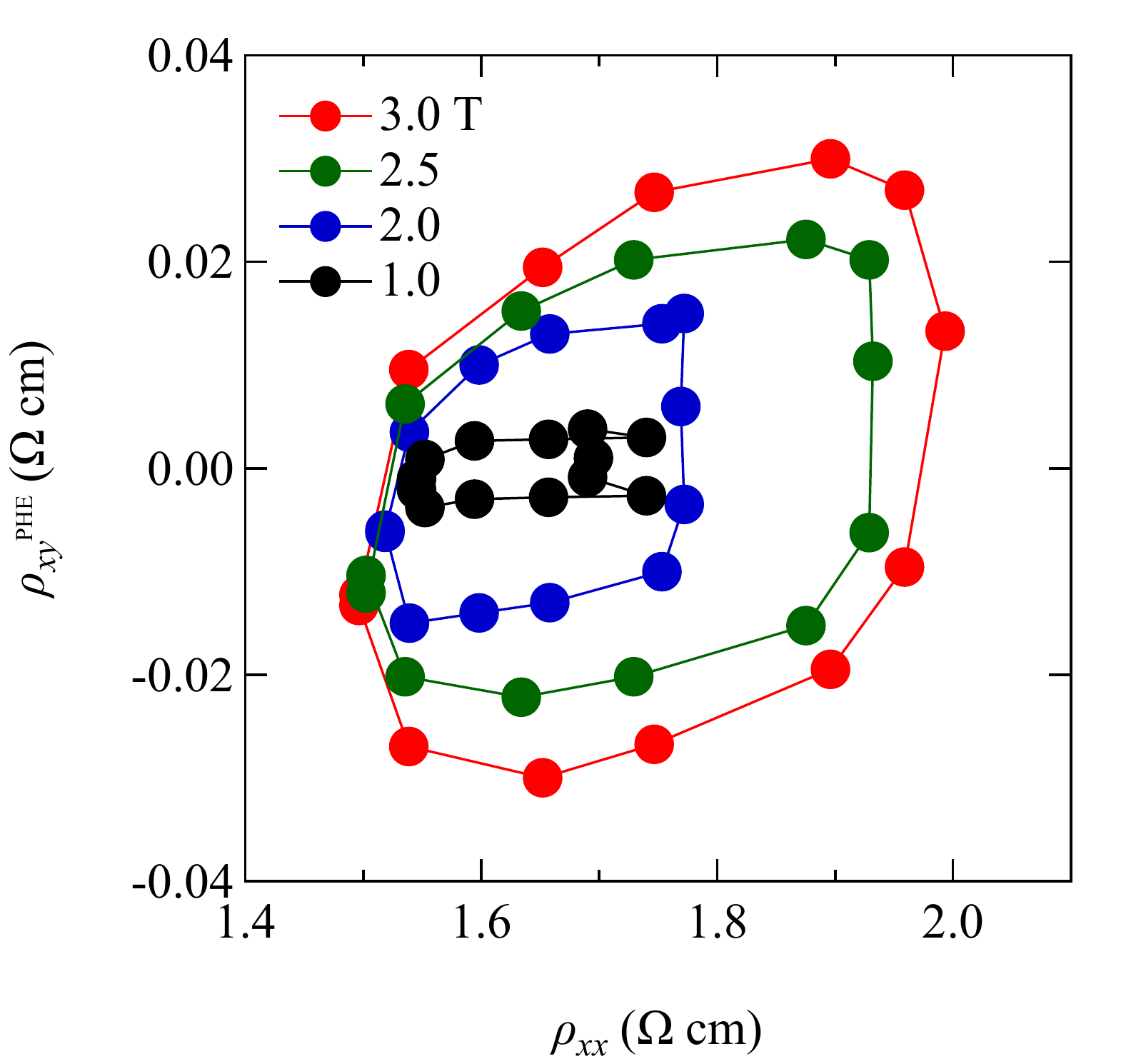}
  \caption{
    \label{fig:parametric}
      Parametric plots of the planar angular magnetoresistance in Sample \#3.
      The orbits at 1.6 K are obtained by plotting
      $\rho_{xx}$ vs. $\rho_{xy}^{\rm PHE}$ with $\phi$ as the parameter
      at magnetic fields 1 T, 2 T, 2.5 T, and 3 T.
    }
\end{figure}

Here, we examine whether the origin of 
the negative magnetoresistance and planar Hall effect 
is due to chiral anomalies or orbital magnetotransport properties 
from the parametric plots of the planar angular magnetoresistance
shown in Fig.~\ref{fig:parametric}.
If the origin of the planar Hall effect is the orbital magnetotransport properties,
the $\rho_{xx}-\rho_{xy}^{\rm PHE}$ parametric plot
exhibits a ``shock-wave'' pattern\cite{Liu2019}. 
However, the plot shows a concentric circle pattern in the magnetic field below 2 T
that excludes the possibility of orbital effects. 
%-------------------------------------------------------------------
% Figure:
%-------------------------------------------------------------------
%-------------------------------------------------------------------

\begin{figure}[htbp]
  \includegraphics[width=0.85 \linewidth]{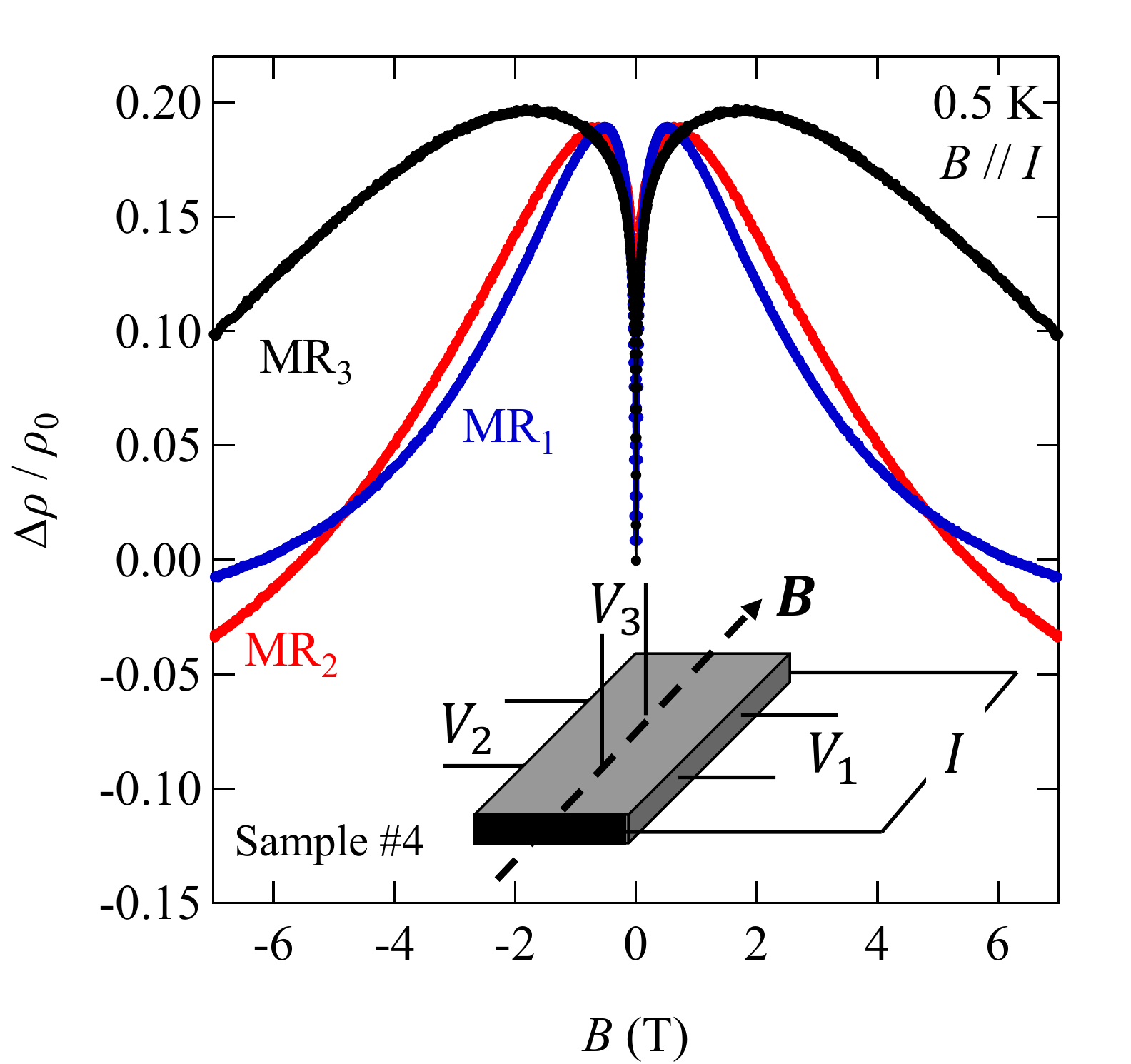}
  \caption{
    \label{fig:mr_sample4}
      Dependence of the longitudinal magnetoresistance (MR$_1$, MR$_2$, and MR$_3$) on the magnetic field for Sample \#4. 
      Measurements were taken at 0.5 K by detecting voltages at the sample edges ($V_1$ and $V_2$) and in the bulk ($V_3$). 
      The measurement configurations are illustrated in the inset.
      }
\end{figure}

    We also examined the current-jetting effect\cite{Reis2016, Ong2021}.
    Our samples have mobilities higher than
    $10^5$~cm$^2$~V$^{-1}$~s$^{-1}$ at low temperatures. 
  In such high-mobility samples, 
  the current-jetting effect can lead to
  longitudinal negative magnetoresistance
  when the current distribution is inhomogeneous.
  When this is the case,
  the magnetoresistance shows qualitatively different behavior
  in bulk and at the sample edges.
  In particular,
  the magnetoresistance is positive in the bulk of the sample.
  In order to examine the current-jetting effect,
  we investigated the longitudinal magnetoresistance
  at the edges (MR$_1$ and MR$_2$) and in the bulk (MR$_3$), 
  as shown in the inset of Fig.~\ref{fig:mr_sample4}.
  Figure ~\ref{fig:mr_sample4} shows that all measurements exhibit
  negative longitudinal magnetoresistance.
  
    If the negative longitudinal magnetoresistance at the edges were primarily due to the current-jetting effect, a positive behavior of the magnetoresistance in the bulk would be anticipated. The observed negative longitudinal magnetoresistance in the bulk, therefore, strongly indicates that the primary cause in our system is not the current-jetting effect.
     
  However, we note that the current-jetting effect may not be completely suppressed 
  because quantitative differences exist between the magnetoresistance 
  at the edges (MR$_1$ and MR$_2$) and in bulk (MR$_3$).

%-------------------------------------------------------------------
% Figure:
%-------------------------------------------------------------------
%-------------------------------------------------------------------

In the following, we interpret the planar Hall effect shown in Fig.~\ref{fig:phe_exp} 
within the framework of a 3D Dirac semimetal with broken chiral symmetry.
The conductivity tensor due to chiral anomalies is described 
as $\Delta \sigma_{ij} \propto B_iB_j$
from a semi-classical approach\cite{Nandy2017,Burkov2017}.
Thus, this formula take the form
\begin{equation}
\label{eq:sxx}
\sigma_{xx}=\sigma_0+\Delta \sigma_{xx} {\rm cos^2}\phi,
\end{equation}
\begin{equation}
\label{eq:syx}
\sigma_{yx}^{\rm PHE}=\frac{\Delta \sigma_{yx}}{2} {\rm sin}2\phi,
\end{equation}
for the azimuthal angle $\phi$ in the magnetic field. 
Here $\Delta \sigma_{xx}=\Delta \sigma_{yx}=\Delta \sigma = A_c B^2$
with $A_c$ being a constant
gives the anisotropy conductivity due to chiral anomaly.
Therefore, the observation of $B^2$ dependence of 
the magnetoconductivity (negative magnetoresistance)  
and planar Hall conductivity with double period for $\phi$ 
is evidence for 3D Dirac semimetal with chiral anomalies. 

%-------------------------------------------------------------------
% Figure:
%-------------------------------------------------------------------
\begin{figure}[htbp]
  \includegraphics[width=0.85 \linewidth]{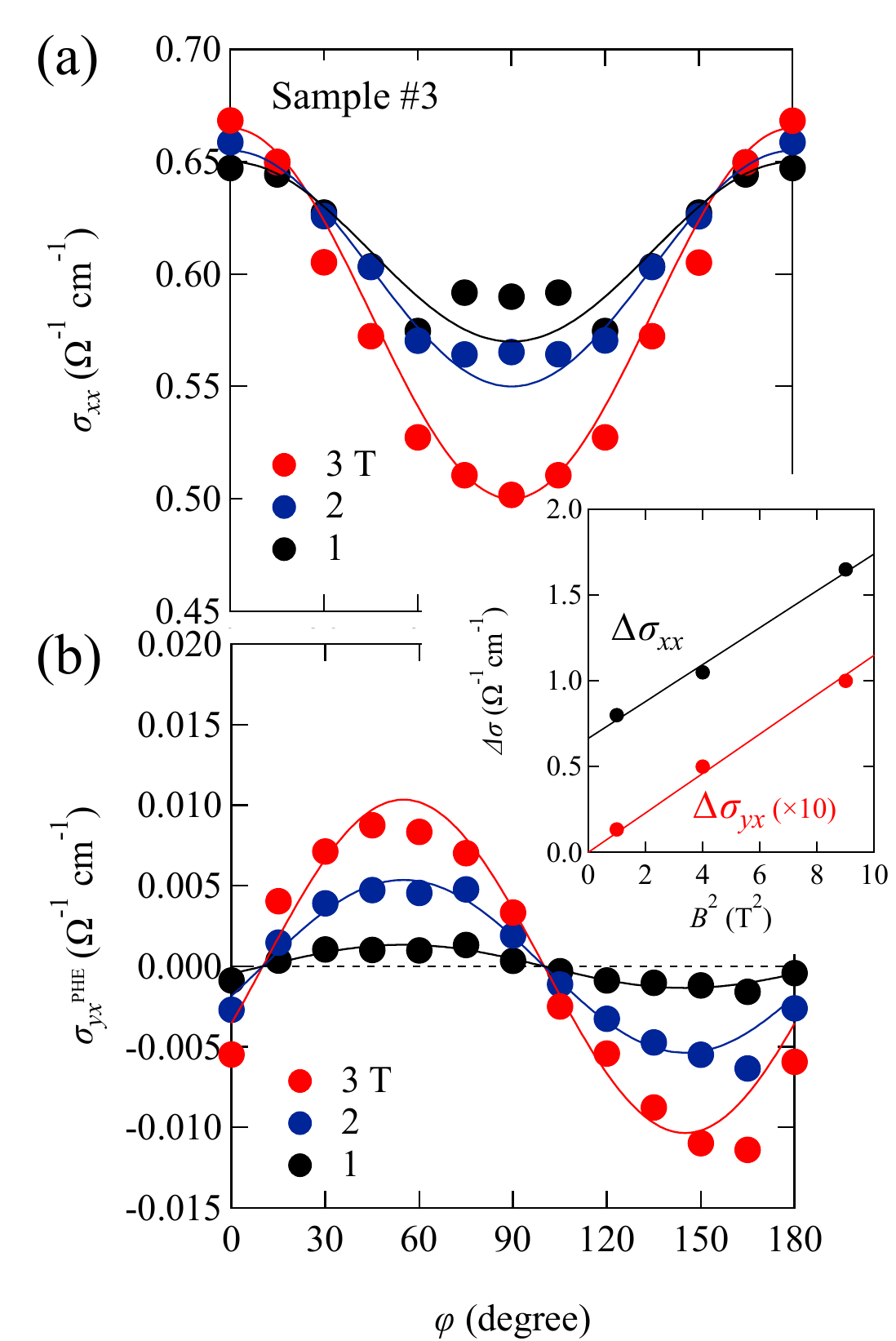}
  \caption{
    \label{fig:phe_sxy}
    $\phi$-dependence of the conductivity (a) and the planar Hall conductivity (b)  at 1 T, 2 T and 3 T. 
    The inset shows the amplitudes of $\phi$-dependence of the conductivity 
    and the planar Hall conductivity as a function of $B^2$.
  }
\end{figure}
%-------------------------------------------------------------------

Now we examine the validity of the equations (\ref{eq:sxx}) and (\ref{eq:syx}).
The conductivity $\sigma_{xx}$ and the planar Hall conductivity $\sigma_{yx}^{\rm PHE}$ 
are calculated as
$\sigma_{xx}=\rho_{xx}/\{\rho_{xx}^2-(\rho_{xy}^{\rm PHE})^2\}$ and 
$\sigma_{yx}^{\rm PHE}=\rho_{xy}^{\rm PHE}/\{\rho_{xx}^2-(\rho_{xy}^{\rm PHE})^2\}$,
respectively.
We show $\sigma_{xx}$ and $\sigma_{yx}^{\rm PHE}$ at 1 T, 2 T and 3 T 
as a function of $\phi$ as shown in Fig.~\ref{fig:phe_sxy}.
The angular dependence of $\sigma_{xx}$
is well described by the formula (\ref{eq:sxx}).
However, the magnetic field dependence is slightly different
from the theoretical prediction.
From the magnetic field dependence of $\Delta \sigma_{xx}$,
we find $\Delta \sigma_{xx}=\Delta \sigma_{0}+A_c B^2$ with 
$\Delta \sigma_{0}\sim 0.07~\Omega^{-1}{\rm cm}^{-1}$ 
and $A_c\sim 0.011~\Omega^{-1}{\rm cm}^{-1}{\rm T}^{-2}$ 
as shown in the inset of Fig.~\ref{fig:phe_exp}(b).
The non-zero value of $\Delta \sigma_{0}$
is associated with the interaction between the electric current 
and the magnetic field for $\phi \neq 0$
and non-chiral states.
On the other hand, these extrinsic effects do not appear in the planar Hall effect.
$\sigma_{yx}^{\rm PHE}$ is well described by Eq.~(\ref{eq:syx})
and we find $\Delta \sigma = 0.001 B^2$
from the analysis of $\sigma_{yx}$.

To conclude, we have observed the hallmarks of the 3D Dirac semimetal state
in \alphaI.
The longitudinal magnetoresistance exhibits a crossover
from a 2D Dirac semimetal to a 3D Dirac semimetal
by decreasing the temperature under high pressure,
and we clearly observed the magnetoresistance that is due to the chiral anomaly.
We have also observed the planar Hall effect.
These results establish that \alphaI is a 3D Dirac semimetal state
and the first organic compound that connects 2D and 3D Dirac semimetal physics.
Since the pressure controls the electronic correlation
and the temperature controls dimensionality,
the system paves the way for investigating the intriguing
interplay between chiral anomaly and spatial dimensionality
and strong electronic correlation.

%------------------------------------------------------------------------
%\begin{acknowledgments}
  %\textbf{Acknowledgments}.
The research was supported by JSPS KAKENHI Grant Numbers 20K03870 and 22K03533.  
%\end{acknowledgments}

\bibliography{../../../../refs/lib}
%\bibliography{lib}
\end{document}